\documentclass[twocolumn,showpacs,preprintnumbers,aps]{revtex4}
\usepackage{graphicx}
\usepackage{dcolumn}
\usepackage{bm}

\begin{document}
\title{Universality in escape from a modulated potential well}

\author{M.I. Dykman, B. Golding, J.R. Kruse, L.I. McCann\onlinecite{byline}, and
D. Ryvkine} \affiliation{Department of Physics and Astronomy,
Michigan State University, East Lansing, MI 48824} \date{\today}
\begin{abstract}
We show that the rate of activated escape $W$ from a periodically
modulated potential displays scaling behavior versus modulation
amplitude $A$.  For adiabatic modulation of an optically trapped
Brownian particle, measurements yield $\ln W\propto (A_{\rm c} -
A)^{\mu}$ with $\mu = 1.5$. The theory gives $\mu=3/2$ in the
adiabatic limit and predicts a crossover to $\mu=2$ scaling as $A$ approaches
the bifurcation point where the metastable state disappears.
%
\end{abstract}


\maketitle

Activated processes, such as escape from a metastable state, are
exponentially sensitive to time-dependent external perturbations.
Escape in driven systems has been investigated in studies of
Josephson junctions \cite{Larkin,Devoret} and infrared photoemission
\cite{photoemission}. It has been also explored recently
in the context of thermally activated magnetization reversal in
nanomagnets driven by a time-dependent magnetic field
\cite{Wernsdorfer97} or a spin-polarized current \cite{Ralph02}.
The strong dependence of the rates of activated processes on the wave
form of the driving field enables selective control of the rates,
which is important for many applications, for example control of rate
and direction of diffusion in spatially periodic systems
\cite{ratchets}.

In this paper we investigate the dependence of the rate of activated
escape $W$ on the amplitude $A$ of a periodic driving field. We show that this
dependence displays universal behavior for large $A$ and investigate
the scaling of $\ln W$ with $A$. Experimental results are
obtained for a colloidal particle trapped in a modulated
optically-generated potential. This system was used earlier
\cite{McCann-99} for a quantitative test of the Kramers theory
\cite{Kramers} of activated escape over a stationary potential
barrier. We observe a power-law behavior of $\ln W$ as $A$ approaches
a critical value. The critical exponent for adiabatic
modulation is $\mu=1.5$, in agreement with theory.  Adiabaticity is
always violated close to the bifurcation point, and we predict the
emergence of a different scaling, with $\mu=2$.

Field dependence of the escape rate is well understood for
adiabatically slow modulation, where the field frequency $\omega_F$ is
small compared to the reciprocal relaxation time $t_{\rm r}^{-1}$. The
driven system remains instantaneously in thermal equilibrium. The
adiabatic escape rate $W_{\rm ad}\propto \exp[-R_{\rm ad}/k_BT]$
depends on the field primarily through the instantaneous activation
energy $R_{\rm ad}(t)$. Even where the modulation of $R_{\rm ad}$ is
small compared to the zero-field $R_{\rm ad}$, it may still
substantially exceed $k_BT$, leading to strong modulation of $W_{\rm
ad}$.

Finding the escape rate becomes more complicated for nonadiabatic
modulation, where $\omega_F\agt t_{\rm r}^{-1}$, because a driven
system is far from thermal equilibrium. The problem has been addressed
theoretically for different models of fluctuating systems
\cite{Larkin,Graham-84,Dykman-97ab,SDG,Hanggi-00,M&S-01}. The
underlying idea is that activated escape results from an optimal
fluctuation, which brings the system from a metastable to an
appropriate saddle-type periodic state. We use it here to draw general
conclusions about the scaling of the period-averaged escape rate $\bar
W$ with the field amplitude $A$.

\begin{figure}
\includegraphics[width=3.0in]{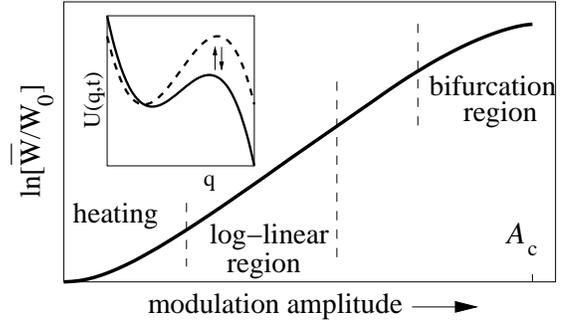}
\caption{Amplitude scaling of the period-averaged rate of escape $\bar
W$ from the modulated potential well shown in the inset ($W_0$ is the
escape rate in the absence of modulation).  The metastable state of
forced vibrations disappears once the modulation amplitude reaches the
critical value ${\cal A}_{\rm c}$.}
%
%
\label{fig:scaling_regions}
\end{figure}

The field dependence of $\ln \bar W$ for $\bar W \ll t_{\rm
r}^{-1},\omega_F$ is sketched in Fig.~\ref{fig:scaling_regions}. For
small $A$, the field is a perturbation. Its major effect is to heat
the system, with the change of the effective energy-dependent
temperature  quadratic in $A$ (the lowest order term after period
averaging). As a result, $\ln\left[\bar W/W_0\right]\propto A^2$,
where $W_0$ is the escape rate in the absence of modulation
\cite{Larkin,Devoret}.  The range of $A$ is limited to $\ln\left[\bar
W/W_0\right]\lesssim 1$.

For stronger modulating fields, the activation energy of escape $R
\propto \ln \bar W$ becomes linear in the amplitude $A$
\cite{Dykman-97ab}. This behavior occurs because the optimal
fluctuation leading to escape is a real-time instanton \cite{Langer},
with duration $\sim t_{\rm r}$. In the absence of modulation, escape
has the same probability to happen at any time. Modulation lifts the
time degeneracy and synchronizes escape events. As a result, even for
zero-mean modulation $\ln\bar W/W_0 \propto A$. For an overdamped
Brownian particle, an explicit solution was obtained \cite{SDG}
throughout the heating and log-linear regions in
Fig.~\ref{fig:scaling_regions}, and the results have been confirmed by
extensive simulations \cite{Chaos-01}.

As we show in this paper, there is another region where the dependence
of $\bar W$ on $A$ is universal and unexpected: the vicinity of the
saddle-node bifurcation point ${\cal A}_{\rm c}$ where the metastable state
of forced vibrations merges with a saddle-type periodic state
\cite{Guckenheimer}. In this range one of the motions of the system
becomes slow.  The universality is related to the corresponding
critical slowing down.  However, it turns out that the system can
display different types of critical behavior depending on the
relationship between $\omega_F$ and the relaxation time in the absence
of driving $t_{{\rm r}0}$.

Experiments on amplitude-dependent escape from a sinusoidally
modulated potential were performed using an
optically-trapped Brownian particle in water.  A configurable
optical potential was constructed by focusing two parallel laser
beams through a single microscope objective lens. Each beam
creates a stable three-dimensional trap as a result of electric
field gradient forces exerted on a transparent dielectric
spherical silica particle of diameter 0.6~$\mu$m.  At the focal
plane, the beams are typically displaced by 0.25 to 0.45~$\mu$m,
creating a double-well potential.  The depths of the
potential wells and the height of the intervening barrier, which is
typically 1 to 10~$k_BT$, are readily controlled by adjusting beam
intensities and separations.

The two HeNe lasers (17~mW, 633~nm) that create the traps are
stabilized by electro-optic modulators and imaged into a sample cell
by a 100x objective.  The beams are mutually incoherent and circularly
polarized when they enter the microscope.  A single trapped colloidal
sphere is imaged onto a digital camera operating at 200 frames/s.  The
$x$ and $y$ coordinates of the sphere's center are computed using a
pattern matching routine that yields spatial resolution better than
±10~nm, whereas the $z$ coordinate parallel to the light wave vector
is extracted by analysis of the image as the particle position
fluctuates about the focal plane.  The overall analysis and storing of
the sphere's $(x,y,z)$ coordinates is accomplished in less than a
frame duration so that no images need to be recorded.

The stability of the system is sufficient to compute the full
3-dimensional optical potential from long-time measurements of the
particle probability density \cite{McCann-99}. The absolute
transition rates calculated from the Kramers expression using
measured curvatures at the stable and unstable points of the
potential are in quantitative agreement with the experimental data
over 4 orders of magnitude.

The electro-optic modulators that stabilize the optical traps also
enable ac modulation with arbitrary waveforms derived from an
electronic signal generator.  When the two beams overlap at the
separations required to form a potential barrier in the range 5 to
8~$k_BT$, each contributes to the position and shape of both
potential minima and the potential in the intermediate range. As a
consequence of this nonlocality, a small modulation of one beam causes
a nearly antisymmetric change of both barrier heights. 

\begin{figure}
\includegraphics[width=3.0in]{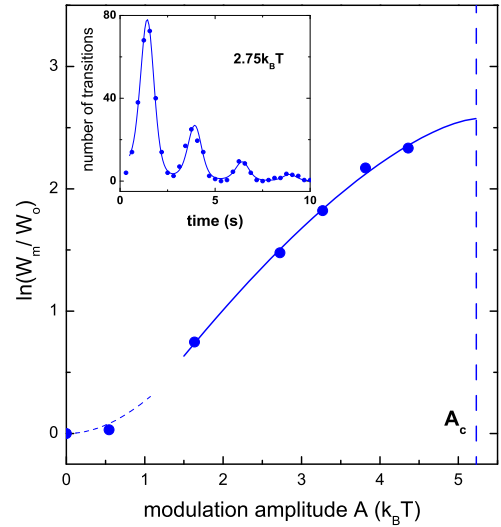}
\caption{Modulation amplitude dependence of the maximal transition
rate $W_m$ normalized to $W_0$, the unmodulated transition rate.  The
data points represent the transition rates for an unmodulated barrier
$\Delta U_0/k_BT = 5.8$ ($W_0 = 0.069$~s$^{-1}$).  The solid line is a
least-squares fit to the data with an asymptotic form described by the
theory (see text).  The modulation frequency $\omega_F/2\pi =
0.4$~Hz. Inset: the number of transitions for $A=2.75 k_BT$ versus
time.}
\label{fig:expt_3_2}
\end{figure}

The results on over-barrier transitions from one of the wells for 
the modulating signal $F(t)=A\cos\omega_Ft$ are
plotted in Fig.~\ref{fig:expt_3_2}. The amplitude $A$ in
Fig.~\ref{fig:expt_3_2} is normalized to $k_BT$, where T = 298~K. The
intrawell relaxation rate in the absence of modulation $t_{{\rm
r}0}^{-1}\approx 10^{3}$~s$^{-1}$ \cite{McCann-99} is much larger than
$\omega_F$, so the data fall into the adiabatic regime. In this case
escape is most likely to occur when the instantaneous potential
barrier $\Delta U(t)$ is minimal. For large modulation amplitude and
for temperatures used here, the maximal escape rate $W_m$ becomes
comparable to the modulation frequency, and therefore it is $W_m$
rather than the period-averaged rate $\bar W$, that is the more
relevant measure. In obtaining $W_m$ we used the distribution of the
dwell time illustrated in the inset to Fig.~\ref{fig:expt_3_2}. The
dwell time was computed for the modulated double-well system as the
time interval between the appearance in, and escape from, a well.

The solid line in Fig.~\ref{fig:expt_3_2} represents a least-squares
fit to the data for large $A$ using the function $\ln W_m/W_0=C_1
-C_2(A_{\rm c}-A)^{\mu}$, where $\mu=1.5$, and the critical modulation
amplitude $A_{\rm c}$ and $C_{1,2}$ are fitting parameters
\cite{uncertainties}.

To interpret these results we will consider activated escape in the
system for $A$ close to the critical amplitude ${\cal A}_{\rm c}$
where the metastable and unstable periodic states merge. In the
adiabatic approximation, the stable and unstable states correspond to
the potential well and the barrier top. Once per period, when the
driving $|F(t)|$ is a maximum (for example when $t=n\tau_F,\,n=0,\pm
1,\ldots$), the stable and unstable states are closest to each other
and the barrier height $\Delta U(t)$ is a minimum. We take $F(0)$
positive, $F(0)=A$. The well and barrier top merge for $A= A_{\rm c}$,
which is the adiabatic bifurcation point.

For small $\delta A=A-A_{\rm c}$ and $|t-n\tau_F|$, the dynamics of
the system near the metastable state is slow \cite{Guckenheimer}. It
is described by the ``soft mode'' with coordinate $q(t)$.  For
negative $\delta A$ the system has a stable and an unstable state,
$q_a$ and $q_b$. We set $q=0$ at the saddle-node state where they
merge in the limit of small $\omega_F$ for $A=A_{\rm c},
t=n\tau_F$. Close to $A_{\rm c}$ and $n\tau_F$ the equation of motion
for $q$, neglecting fluctuations, takes the form
\begin{equation}
\label{mean-field}
t_{{\rm r}0} {dq\over dt} = q^2 + F(t)-A_{\rm c}
\end{equation}
($t_{{\rm r}0}$ is the relaxation time for $F=0$, $\omega_Ft_{{\rm
r}0}\ll 1$). In the adiabatic approximation, $F(t)$ depends
parametrically on $t$. Activated transitions occur over the
instantaneous barrier $\Delta U(t)= (4/3)[A_{\rm
c}-F(t)]^{3/2}$.

The relaxation time of the system $t_{\rm r}=t_{{\rm r}0}[
A_{\rm c}-F(t)]^{-1/2}$ becomes large close to the bifurcation
point. Therefore even a slowly varying field will ultimately become
too fast for the system to follow adiabatically.  To allow for
nonadiabaticity we expand $F(t)$ in the region of small
$\omega_F|t-n\tau_F|$. The equation of motion (\ref{mean-field}) in
scaled variables $Q=q/\Omega,\; \tau=\Omega (t-n\tau_F)/t_{{\rm r}0}$
is then
\begin{eqnarray}
\label{nonad_eom}
\dot Q =  K(Q,\tau), \; K= Q^2 + \zeta - \tau^2,\;
\end{eqnarray}
with $\zeta= \Omega^{-2}(A-A_{\rm c})$. In
Eq.~(\ref{nonad_eom}) $\dot Q\equiv dQ/d\tau$, and $\Omega =
(-{1\over 2}t_{{\rm r}0}^2 d^2F/dt^2)^{1/4}$ (the derivative is
evaluated at $t=0$ and $A=A_{\rm c}$). For sinusoidal
modulation $\Omega = (A_{\rm c}t_{{\rm
r}0}^2\omega_F^2/2)^{1/4}$.

\begin{figure}
\includegraphics[width=2.0in]{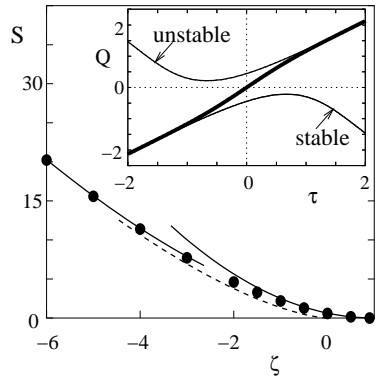}
\caption{The scaled activation energy of escape $S\approx -\tilde D\ln
W_t$
(dots), obtained numerically from the variational problem
(\protect\ref{act_energy}), vs. $\zeta=\Omega^{-2}(A-A_{\rm c})$. In
the adiabatic approximation $S={4\over 3}(-\zeta)^{3/2}$ (the dashed
line), which explains the results in
Fig.~\protect\ref{fig:expt_3_2}. The nonadiabatic bifurcation point is
$\zeta = 1$. The solid lines show the corrected adiabatic expression
for large-$|\zeta|$, Eq.~(\protect\ref{large_zeta}), and the
nonadiabatic behavior $S\propto (1-\zeta)^2$,
Eq.~(\protect\ref{true_bif}). Inset: the attracting and repelling
states $Q_a$ and $Q_b$ (thin lines) for weakly nonadiabatic driving,
Eq.~(\protect\ref{nonad_eom}) with $\zeta = 0.95$, and the most
probable escape path (thick line).}
\label{fig:true_bif_states}
\end{figure}

The dynamics (\ref{nonad_eom}) is determined by one dimensionless
parameter $\zeta$. For $-\zeta\gg 1$, the adiabatic approximation
applies, and the stable and unstable states are given by
$Q_{a,b}(\tau)=\mp(-\zeta+\tau^2)^{1/2}$.
On the other hand, for $|\zeta|\lesssim 1$ the time dependence of the
states becomes distorted and asymmetric, cf. inset in
Fig.~\ref{fig:true_bif_states}.  For $\zeta=1$, the states given by
Eq.~(\ref{nonad_eom}) merge and $Q_a(\tau)=Q_b(\tau) =\tau$. The
corresponding nonadiabatic bifurcational value of the amplitude is
$A_{\rm b} = A_{\rm c}+\Omega^2$.

Escape from the attractor can be quite generally described by adding a
random Gaussian force to Eq.~(\ref{mean-field}). Because the system
dynamics is slow, this force is effectively $\delta$-correlated (white
noise), with intensity $D\propto k_BT$ for thermal noise. The
corresponding force $\xi(\tau)$ in Eq.~(\ref{nonad_eom}) for $\dot Q$
has a correlator
$\langle\xi(\tau)\xi(\tau')\rangle =
2\tilde D\delta (\tau-\tau'), \; \tilde D=
D/\Omega^3t_{{\rm r}0}.$

The noise-driven system is most likely to escape during the time when
$F(t)$ is close to its maximum. The total probability to escape over
this time is $W_t=\bar W\tau_F$, where $\bar W$ is the period-averaged
escape rate. We will calculate $W_t$ assuming that the noise intensity
$D$ is the smallest parameter of the theory, in which case $W_t\propto
\exp(-R/D)$. The calculation can be done by generalizing the instanton
technique \cite{Chaos-01} to the Langevin equation
$\dot Q= K(Q,\tau) + \xi(\tau)$. The activation energy of escape
$R=(D/\tilde D)S$ is given by the solution of the variational problem
\begin{equation}
\label{act_energy}
S={1\over 4}\min\int\nolimits_{-\infty}^{\infty}d\tau (\dot Q - K)^2.
\end{equation}
The boundary conditions are $Q(\tau)\to Q_a(\tau)$ for $\tau\to
-\infty$ and $Q(\tau)\to Q_b(\tau)$ for $\tau \to \infty$. They
correspond to the picture in which, prior to escape, the system is
performing small fluctuations about the metastable state $Q_a$. Escape
occurs as a result of a large fluctuation that brings the system to
the saddle state $Q_b$. The most probable among such fluctuations is
the one in which the system moves along a trajectory [the most
probable escape path (MPEP)] that minimizes the functional
(\ref{act_energy}).  From Eq.~(\ref{act_energy}), the activation
energy $R$ is a function of one parameter $\zeta$, shown in
Fig.~\ref{fig:true_bif_states}.

An explicit expression for $R$ can be
obtained in the adiabatic limit $-\zeta \gg 1$. Here, one can ignore
the term $\tau^2$ in $K$ (\ref{nonad_eom}) for typical $|\tau|\lesssim
\zeta^{-1/2}$. Then the MPEP is an instanton
$Q_{\rm opt}(\tau,\tau_0) =|\zeta|^{1/2}\tanh\left[|\zeta|^{1/2}(\tau -
\tau_0)\right]$
centered at an arbitrary $\tau_0$. The adiabatic activation energy is
$R\propto S=(4/3)(-\zeta)^{3/2}\propto (A_{\rm c}-A)^{3/2}$
\cite{3/2}.  This agrees with the experimental data shown in
Fig.~\ref{fig:expt_3_2}.

The time-dependent term in $K$ lifts translational invariance of
the instanton $Q_{\rm opt}(\tau,\tau_0)$ and synchronizes escape
events. The action $S$ is minimal for $\tau_0=0$. To lowest order
in $1/\zeta$
\begin{equation}
\label{large_zeta}
S\approx (4/3)|\zeta|^{3/2} + (\pi^2/6)|\zeta|^{-1/2}.
\end{equation}
From (\ref{large_zeta}), the nonadiabatic correction to the activation
energy diverges as $|\delta A|^{-1/2}$ as $A$ approaches $A_{\rm c}$.

The activation energy may also be found explicitly close to the
bifurcation point $\zeta = 1$. In this case escape is most likely to
occur while the coexisting attracting and repelling trajectories
$Q_a(\tau)$ and $Q_b(\tau)$ stay close to each other.
Because of the special structure of the states $Q_{a,b}(\tau)$, the
variational problem (\ref{act_energy}) for the MPEP can be linearized and
solved, giving
\begin{equation}
\label{true_bif}
S =(\pi/8)^{1/2}(1-\zeta)^2,\; 1-\zeta \ll 1.
\end{equation}

Eq.~(\ref{true_bif}) shows that, near the nonadiabatic bifurcation
point $A_{\rm b}$, there emerges another scaling region, where the
activation energy $R\propto (A_{\rm b}-A)^2$. The nonadiabatic
exponent is thus $\mu=2$, in contrast to $\mu= 3/2$ in the adiabatic
limit. It describes $R(A)$ for the modulation amplitude lying between
$A_{\rm c}$ and $A_{\rm b}$, as seen from
Fig.~\ref{fig:true_bif_states}. The width of this interval is $\propto
\omega_F$ provided $\omega_Ft_{{\rm r}0}\ll 1$.

The above theory does not apply in the exponentially narrow range
$1-\zeta \lesssim \exp[-C/\omega_Ft_{{\rm r}0}]$ ($C\sim 1$). This is
because, as $A$ approaches $A_{\rm b}$, the relaxation time of the
system $t_{\rm r}$ becomes logarithmically long. Then the expansion of
the driving force in $t/\tau_F$ that leads to the equation of motion
(\ref{nonad_eom}) becomes inapplicable.  For $\omega_Ft_{\rm r}\gg 1$
the duration of the MPEP greatly exceeds the period $\tau_F$. In this
case the activation energy again displays the $\mu=3/2$ power-law
dependence on the distance to the true bifurcation value of the
amplitude ${\cal A}_{\rm c}$. The $({\cal A}_{\rm c}-A)^{3/2}$-law
also applies asymptotically if the driving is nonadiabatic for small
$A$.

The experimental data discussed above refer to very small
$\omega_Ft_{{\rm r}0} < 10^{-3}$. Therefore the nonadiabatic
region of modulating amplitudes is narrow and the activation
energy there is less than $k_BT$. Higher modulation frequencies
will be required to detect the crossover from the observed
$\mu=3/2$ scaling to the nonadiabatic $\mu=2$ scaling. It appears
that the $\mu=3/2$ scaling function describes the data in a broad
range of $A$. It thus provides a good interpolation of the
activation energy from the linear in $A$ to the
critical region.

In conclusion, we have identified several regions where the activation
energy $R$ of escape from a metastable potential displays scaling
behavior as a function of the amplitude $A$ of an externally applied
periodic modulation. We have demonstrated with a system whose
potential was directly measured that, near the amplitude $A_{\rm c}$
where the local minimum and maximum of the potential contact one
another, $R\propto (A_{\rm c}-A)^{3/2}$. We have also shown that,
because of nonadiabaticity near the bifurcation point, there
necessarily emerges a region where the scaling exponent changes from
$\mu=3/2$ to $\mu=2$. We expect that these scalings can be observed in
other systems, such as modulated Josephson junctions \cite{Devoret}
and nanomagnets \cite{Wernsdorfer97,Ralph02,Koch00}. The ideas discussed here
can be used to gain additional insight into the physics of these
systems and more generally, activated effects in nonequilibrium
systems.

This research was supported by the NSF DMR-9971537 and NSF
PHY-0071059.


\begin{thebibliography}{90}
\bibitem{uncertainties} The results of the least squares fit in
Fig.~\protect\ref{fig:expt_3_2} are: $A_{\rm c}/k_BT =5.2\pm 0.7,
\,C_1= 2.6\pm 0.3, C_2=0.27\pm 0.05$. We also allowed the critical
exponent $\mu$ to vary with a four-parameter fit, which led to the
uncertainty $\delta\mu=\pm 0.1$.
\end{thebibliography}
\end{document}